\documentclass[amsmath, amssymb,10pt, aps, prb, notitlepage, longbibliography, superscriptaddress]{revtex4-2}
\usepackage{graphicx}
\usepackage{amsmath}
\usepackage{hyperref}
\usepackage{xcolor}

\newcommand{\cS}{\mathcal{S}}
\newcommand{\hSz}{\hat{S}^{z}}
\newcommand{\hSx}{\hat{S}^{x}}
\newcommand{\hSy}{\hat{S}^{y}}

\newcommand{\bh}{{\bf h}}

\newcommand{\bk}{{\bf k}}

\newcommand{\br}{{\bf r}}

\def\bnabla{{\bm \nabla}}

\newcommand{\hb}{{\hat b}}

\newcommand{\hn}{{\hat n}}

\newcommand{\hS}{{\hat S}}

\newcommand{\cH}{\hat{\cal H}}

\makeatletter
\newcommand{\citeordered}[1]{%
  \begingroup
  \def\NAT@sort{0}%
  \def\NAT@cmprs{0}%
  \cite{#1}%
  \endgroup
}
\makeatother

\begin{document}

\title{Why ``quantum spin liquids'' are quantum singlet liquids}


\author{Sergey Syzranov}
\author{Arthur P. Ramirez}
\affiliation{Physics Department, University of California Santa Cruz, Santa Cruz, California 95064, USA}

\begin{abstract}
   The last couple of decades have seen enormous theoretical and experimental research efforts aimed at finding quantum spin liquids (QSLs) in geometrically frustrated (GF) magnetic materials. 
   Experimental demonstrations of QSLs are challenging due to the indirect character of possible QSL signatures and the presence of impurities in materials, which may destroy or obscure a QSL. At the same time, the theoretical literature often lacks consensus about the ground states of quantum spins on even the simplest GF magnetic lattices. 
   We summarize experimental evidence that in the impurity-free limit, available GF magnetic materials that do not develop magnetic order at low temperatures would form
   {\it a quantum singlet liquid}, i.e. a disordered 
   state of non-magnetic excitations. We further confirm with theoretical arguments the existence of such a singlet liquid and its gapless character in materials {with half-integer spins per primitive magnetic cell}.
   Such a liquid manifests itself in the thermodynamics of GF magnets while having no magnetic features at low temperatures.
\end{abstract}

\maketitle

Since Anderson's 1973 proposal~\cite{Anderson:RVB} of a liquid-like ground state of spin systems, immense global efforts have been aimed and finding and conclusively demonstrating quantum spin liquids (QSLs) in specific materials. An indispensable attribute of a spin liquid is the lack of long-range magnetic order. To constitute a {\it quantum} spin liquid, the material is also usually required to have unusual quantum properties, such as topological features, fractionalized excitations, or long-range entanglement, although there is no universal, commonly accepted definition of a QSL.

Despite the significant research efforts, there is no consensus about whether a QSL has been realized in any particular material.
One of the obstacles for decisive realization/demonstration of QSLs is quenched disorder: even weak randomness induces a spin-glass transition~\cite{Villain:insulatingspinglasses,SaundersChalker:spinfreezing,AndreanovSherrington:spinglass},
as observed since the earliest studies~\cite{GreedanRazavi:spinglass,RamirezCooper:frustration,SyzranovRamirez:eminuscent,RamirezSyzranov:hiddenscale} in geometrically frustrated magnetic materials,
with the spin-glass state displacing a QSL and mimicking some of its expected features~\cite{KimchiSenthil:randomsinglets,ZhuChernyshev:mimicry,Norman:herbertsmithite}.
While it is commonly hoped that
a QSL can be approached by purifying the material, there is little understanding of the appropriate purity requirements.
Furthermore, for achievable purity levels, reducing the number 
of non-magnetic impurities has been shown to favor spin-glass freezing incompatible with a QSL~\cite{SyzranovRamirez:eminuscent,Syzranov:vacancies}.
Another obstacle is the indirect character
of potential experimental signatures of QSLs: although such signatures, e.g. the temperature dependence of the heat capacity, may be consistent with particular QSL models, they can also admit alternative explanations (see, e.g., Refs.~\citeordered{Wen:QSLidentificationReview,Loidl:KitaevPossibleEvidence,RamirezSyzranov:hiddenscale}).

On the theoretical side, tremendous numerical and analytical efforts have been directed at identifying the QSL states specific to particular geometries of magnetic lattices and predicting manifestations of those states. 
To identify possible phases of very clean magnets, numerous analytical methods were invented, such as mean-field techniques in the parton representation~\cite{AffleckMarston:largeN,ArovasAuerbach:Schwinger,ReadSachdev:largeN,Wen:Z2meanfield,Wen:quantumorders,SavaryBalents:QSLreview}, large-$S$ expansions~\cite{Haldane:largeS,Chubukov:orderfromdisorder,ChubukovSenthil:largeS}, and phenomenological approaches~\cite{HermeleBalents:pyrochlorephotons,SenthilFisher:Z2}.
Only a small number of QSL models,
typically where both the form and strength of interactions are fine-tuned, allow for exact analytical results~\cite{Kitaev:honeycomb,ShastrySutherland,RK,ChanglaniFradkin:exactpoint}.

Methods other than exact solutions are typically uncontrolled and often lead to rather uncertain predictions. For example, for the Heisenberg model on the kagome lattice, one of the most thoroughly investigated potential QSL platforms, various theoretical methods have pointed to a
variety of drastically different possible ground states:
a Dirac QSL
\cite{RanWen:kagome,HePollmann:kagomeDirac,HuXiang:spectralfunction}, a
$\mathbb{Z}_2$ gapped QSL~\cite{Sachdev:QMC,YanWhite:kagome,DepenbrockMcCullochSchollwock:DMRG},
a chiral spin liquid
\cite{SunZhou:CSLkagome,HeChen:CSLkagome},
a valence-bond crystal~\cite{SinghHuse:VBC},
and other states (see, e.g., Refs.~\cite{RyuFisher:algebraicvortex,DuricSengupta:spinonPDW,KamalYao:comment,GongSheng:globalphasediagram}). Similar uncertainties persist for possible quantum spin states on other common geometrically frustrating (GF) lattices.

The purpose of this paper is to point out a generic unifying trend among the lowest-energy states of disorder-free GF magnets, the largest set of candidate materials that may host QSLs. We demonstrate that 
{\color{purple}\bf the lowest-energy states of experimentally available GF materials form a {\it\bf quantum singlet liquid}, i.e. a non-ordered ensemble of gapless non-magnetic excitations}.
Such liquids manifest themselves in the thermodynamics of GF materials without having magnetic signatures at lowest temperatures. The emergence of quantum singlet liquids
is also consistent with the observed lack of magnetic field dependence of the low-temperature specific heat in GF materials ~\cite{RamirezWinklemann:singlets,NakatsujiMaeno:spindisorder}.

Although the {\it quantum singlet liquid} may not, on its own, uniquely identify the ground state on certain particular lattices (for example, a low-energy gapless $U(1)$ QSL conjectured to exist on the pyrochlore lattice~\cite{HermeleBalents:pyrochlorephotons} and quantum dimer QSLs would constitute singlet liquids), it 
should be used as a basis for constructing possible 
ground states of GF materials. The concept of a quantum singlet liquid can also be used to further model experimental properties, e.g. heat transport, in such materials.


{\bf\color{blue} Experimental evidence.}
Impurity-free magnetic properties of GF materials are expected to have one characteristic energy scale: the Curie-Weiss temperature $\theta_\text{CW}$~\cite{Blundell:book,Ramirez:GFreview}. 
The thermodynamics of a typical GF magnet, however, generically exhibit an additional, lower energy scale $T^*$, which we dubbed 
the ``hidden energy'' scale in Ref.~\cite{RamirezSyzranov:hiddenscale}.

Both the Curie-Weiss temperature $\theta_\text{CW}$
and the hidden energy scale $T^*$ manifest themselves as peaks in heat capacity $C(T)$~\cite{PoppSyzranov:fratio}.
Some materials exhibit two distinct, well-separated $C(T)$ peaks at temperatures $T^*$ and $\theta_\text{CW}$. In other materials, the peaks may partially merge, leading to the formation of a $C(T)$ ``shoulder'' at the temperature $T^*$~\cite{SugiuraShimizu:KHAF,Munehisa:KHAF,ChenQu:THAF,PrelovsekKokalj:THAF,SchnackRichter:kagome,Gonzalez:THAF}.
All relevant neutron scattering
and specific heat data also show a buildup of short-range-order (SRO)
among spins at the temperature $T^*$~\cite{RamirezSyzranov:hiddenscale,Popp:neutrons}. 
In the presence of randomly located impurities,
$T^*$ often determines the spin-glass-freezing temperature~\cite{SyzranovRamirez:eminuscent}.

The absence of magnetic features at temperatures $T\lesssim T^*$ indicates that the lowest-energy excitations in impurity-free magnetic materials are non-magnetic, hereafter referred to as singlets. The available thermodynamics and magnetic data show, therefore, the non-magnetic character of lowest-energy disordered states of GF 
magnetic materials. The ``quantum spin liquids'' that may emerge in sufficiently clean GF magnetic materials are, therefore, quantum singlet liquids.
The singlet character of such a state is further reinforced by the lack of the magnetic-field dependence of the heat capacity at temperatures $T\lesssim T^*$ in materials where such measurements are available~\cite{RamirezWinklemann:singlets,NakatsujiMaeno:spindisorder}.




{\bf\color{blue} Boson representation.}
The emergence of the quantum singlet liquid at low temperatures can be naturally understood by mapping the spin degrees of freedom to those of bosonic particles.
Consider, for concreteness, a spin-$1/2$ Heisenberg XXZ model described by the Hamiltonian
\begin{equation}
\cH_\text{XXZ} =  \sum_{(ij)} \left( J^\parallel_{ij}\hSz_i \hSz_j + J^\bot_{ij} \hSx_i \hSx_j + J^\bot_{ij} \hSy_i \hSy_j \right)
-\sum_i \bh\cdot{\bf\hS}_i,
\label{Hspin}
\end{equation}
where the summation is carried out over all pairs $(ij)$ of spin sites with interacting spins; $J^\parallel_{ij}$ and $J^\bot_{ij}$ are the antiferromagnetic exchange couplings that describe the interactions between the spins at sites $i$ and $j$; and $\bh$ is an external magnetic field.
In what follows, we assume that the model under consideration has a unique ground state; in the opposite case, the ground-state degeneracy can be lifted by an infinitesimal perturbation.

This model can be mapped to the the Bose-Hubbard model 
described by the Hamiltonian
\begin{align}
    \cH_\text{boson}=
    &U\sum_i\left(\hn_i-\frac{1}{2}\right)^2+\sum_{(ij)}J^\parallel_{ij}\left(\hn_i-\frac12\right)\left(\hn_j-\frac12\right)
    +\frac12\sum_{(ij)} J^\bot_{ij}
    \left(\hb_i^\dagger \hb_j+\hb_j^\dagger \hb_i\right)
    \nonumber \\
    &-\sum_i h_z\left(\hn_i-\frac12\right)-\frac12\sum_i \left(h^\dagger\hb_i+h^-\hb^+_i\right);
    \qquad \hn_i=\hb_i^\dagger \hb_i , \quad
    h^\pm=h_x\pm ih_y, \quad
       U \rightarrow+\infty
    \label{Hboson}
\end{align}
using variable correspondence rules
\begin{align}
    \hb_i \leftrightarrow \hS_i, 
    \quad
        \hb_i^\dagger \leftrightarrow \hS_i^\dagger,
    \quad
    \hn_i-\frac{1}{2}\leftrightarrow \hS_z,
\end{align}
where $\hb_i$ and $\hb_i^\dagger$ are the boson annihilation and creation operators; and $\hn_i$ is the boson number at site $i$. 
Although we describe the spin-boson mapping for spin-$1/2$ systems, it can be readily generalized to higher spins.

A 3D translationally invariant system of bosonic particles is generically expected to lack the metallic phase~\cite{Efetov:SIT,FisherFisher:bosonlocalization,FisherGrinsteinGirvin:UniversalConductance,Phillips:BoseMetalElusive,CapogrossoSansone:SIT}:
the system is either insulating, with gapped {\it monopole} excitations or superfluid, with the superfluid order parameter $\left<\hat b_i\right>\neq 0$. In terms of the equivalent spin degrees of freedom (see Table~\ref{TableSpinBosonCorrespondence}), 
for the XXZ model with the Hamiltonian~\eqref{Hspin},
the absence of the bosonic metallic phase translates into either having transverse order or magnetic excitations being gapped.


\renewcommand{\arraystretch}{1.25}

\begin{table}[t!]

\begin{center}
\begin{tabular}{|p{0.45\textwidth}|p{0.45\textwidth}|}
\hline
\textbf{Boson phase} & \textbf{Spin counterpart} \\
\hline

\textbf{Superfluid}\par
$\langle \hat b_i\rangle\neq0$; no density wave
&
\textbf{XY-ordered magnet}\par
$\langle \hat S_i^+\rangle\neq0$; no periodic $\langle \hat S_i^z\rangle$
\\
\hline

\textbf{Supersolid}\par
$\langle \hat b_i\rangle\neq0$; density wave
&
\textbf{XY + longitudinal order}\par
$\langle \hat S_i^+\rangle\neq0$; periodic $\langle \hat S_i^z\rangle$
\\
\hline

\textbf{Uniform Mott insulator}\par
$\langle \hat b_i\rangle=0$; no density wave; particle/hole gap $>0$
&
\textbf{Spin liquid}\par
$\langle \hat S_i^+\rangle=0$; no periodic $\langle \hat S_i^z\rangle$\par
$S^z$-carrying excitations gapped
\\
\hline

\textbf{Density-wave Mott insulator}\par
$\langle \hat b_i\rangle=0$; density wave; particle/hole gap $>0$
&
\textbf{Longitudinally ordered magnet}\par
$\langle \hat S_i^+\rangle=0$; periodic $\langle \hat S_i^z\rangle$\par
$S^z$-carrying excitations gapped
\\
\hline

\textbf{Bose metal}\par
$\langle \hat b_i\rangle=0$; no density wave; compressibility $\neq0$
&
\textbf{Compressible spin liquid}\par
$\langle \hat S_i^+\rangle=0$; no periodic $\langle \hat S_i^z\rangle$;
$\chi_{zz}\neq0$
\\
\hline

\end{tabular}
\end{center}

\caption{Correspondence between the phases in the XXZ spin model [see Hamiltonian~\eqref{Hspin}] and the equivalent Bose-Hubbard model [see Hamiltonian~\eqref{Hboson}].
\label{TableSpinBosonCorrespondence}
}

\end{table}


Therefore, a QSL, in which 
long-range magnetic order is absent,
cannot host gapless {\it magnetic} excitations. At the same time, Lieb-Shultz-Mattis-type theorems~\cite{LiebSchultzMattis,Oshikawa:commensurability,Hastings:LSM} disallow gapped excitation spectra in systems with unique ground states {and half-integer spins per primitive cell}. Such a liquid, therefore, must host gappless non-magnetic excitations (singlets).
Thus, in agreement with the experimental phenomenology,
a quantum singlet liquid emerges in such a material at low temperatures.

{\bf\color{blue} Excitation field theories.}
The described behavior in GF materials is consistent with the low-energy magnetic excitations with the magnetization  [$U(1)$ spin charge] $m$ described by the effective field theory with the Matsubara action
\begin{align}
    \cS= \int_0^\beta dt\int_\br \left[\bar{\psi}\left(\partial_\tau  +\frac{\hat \bk^2}{2M}+r\right)\psi+
   g_\bot\left( h^+\right)^m\psi+
   g_\bot\bar{\psi} \left(h^-\right)^m+
   g_\parallel h_z\bar{\psi}\psi+\ldots\right],
   \label{MagnonAction}
\end{align}
where the complex numbers $\bar{\psi}$ and $\psi$ are the bosonic variables; the parameter  $r$ mimics the magnon energy gap in the ``spin liquid'' (insulating bosonic) phase if $r>0$; $r<0$ corresponds to the XY-ordered magnetic (superfluid) phase; $\hat \bk =-\hbar\partial_\br$; $\frac{\hat \bk^2}{2M}+r$ is the magnon dispersion near the minimum; $g_\parallel$ and $g_\bot$ are constants; and $\ldots$ represents the higher-order terms in $\psi$ and $\bar{\psi}$. Hereafter, we assumed the dispersion
to be isotropic for simplicity.
The $h^\pm$, $h^z$-dependent source terms in the action~\eqref{MagnonAction} describe the magnon coupling to the magnetic field. 

By contrast, the non-magnetic excitations, which dominate the low-temperature thermodymics of GF materials, are $U(1)$ neutral and are described by the action
\begin{align}
    \cS_\text{singlet}= \int_0^\beta dt\int_\br \left[
    \left(\partial_\tau \phi\right)^2
    +a\left(\bnabla \phi\right)^2+\ldots
    \right]
  \label{SingletAction}
\end{align}
with the real field $\phi$, where $a>0$ is a constant.
In such a theory, there is no excitation coupling to the magnetic field. 
Equation~\eqref{SingletAction} shows symmetry-allowed terms of the lowest-order in the gradients of the Matsubara time $\tau$ and the spatial coordinates $\br$, which, if non-vanishing, lead to linearly dispersive excitations.
Quadratic temperature dependencies of the specific heat, consistent with linearly dispersive excitations, are observed in certain layered GF magnets~\cite{RamirezSyzranov:hiddenscale}. 
The microscopic origin of that temperature dependence, however, requires further investigation. Linearly dispersive singlet excitations, ``lattice photons'', are also commonly believed to 
emerge on certain lattices, such as the pyrochlore lattice and corner-sharing octahedra~\cite{HermeleBalents:pyrochlorephotons,Shannon:QMCQED,SavaryBalents:QSLreview}.


{\bf\color{blue} Discussion.} 
In our view, the importance of understanding the nature of the ground states of GF quantum spin systems, as well as the common pitfalls of many existing theoretical and numerical approaches (e.g. the parton construction) to these systems, is aptly captured by Y.~Yao, C.J.~Umrigar, and V.~Elser's metaphor in
Ref.~\cite{YaoElser:kagomechemistry}:

\vspace{1mm}
\noindent
{\it \color{purple}
In condensed-matter phenomena, we are guided by the
Landau paradigm, where the low-energy physics is derived
from general characteristics of the ground state. This strategy, while enormously successful, assumes we have a firm grasp of the ``chemistry'' of our system. To see what can go wrong, consider the case of the quantum chemist who undertakes a study of hydrogen oxides, in particular, the contentious $2–1$ compound. One is limited to studies of small clusters and is frustrated because the ground-state properties (structure
factor, phonon spectrum, etc.) depend sensitively and
unpredictably on system size, boundary conditions, pressure.
By going straight to the lowest energy properties of the
system, the researcher has failed to notice that the atoms single-mindedly first form $H_2O$ molecules, and it is the quirky
interactions among these constituents that is responsible for the complex behavior of the bulk compound.
}
\vspace{1mm}

In this work, we aim to elucidate a generic analogue of such ``$H_2O$ emergence'' across a broad class of experimentally accessible GF magnetic materials by identifying a common basis for their low-energy states and thus facilitating realistic predictions of such states in the quenched-disorder-free limit.

We argue that these materials generically develop a quantum singlet liquid, i.e. a disordered state of non-magnetic excitations. This liquid generically manifests itself in the thermodynamics and neutron scattering at a characteristic temperature $T^*$ below the Curie-Weiss temperature  $\theta_\text{CW}$. While the existence of such a liquid does not uniquely specify the ground state - depending on the material, it could be, e.g., lowest-energy excitations in $U(1)$ gapless QSL~\cite{HermeleBalents:pyrochlorephotons} or a quantum-dimer liquid -
it constitutes a basis on which the actual ground states can be  constructed.

{Essential to further exploration of possible ground states is the effect of quenched disorder on the quantum singlet liquid. As the theoretical arguments provided above focus on systems  with half-integer spins per magnetic primitive cell, further analytical and numerical studies are required for systems with integer spins per cell.}



{\bf\color{blue} Acknowledgements.}
We are indebted to P.~Popp, S.~Rozenkranz, M.~Sedik, S.~Sun, S.~Zhu for collaboration on related subjects. 
Our work has been supported by the NSF grant DMR2218130.
In preparation of this manuscript,
ChatGPT-5.6 and Claude Fable 5.1 have been used for testing arguments, assistance in reviewing literature, improving presentation, and summarizing commonly known facts.
All the arguments and results come from the authors, and all AI outputs have been reviewed and verified by the authors. The authors take full responsibility for the contents of the manuscript. 


%









\end{document}